**Responsible AI for Test Equity and Quality:**

**The Duolingo English Test as a Case Study**

Jill Burstein[a,1], Geoffrey T. LaFlair[a], Kevin Yancey[a], and Alina A. von Davier[a]

Duolingo, Inc.

5900 Penn Avenue

Pittsburgh, PA 15206

{jill, geoff, kevin, avondavier}@duolingo.com

Ravit Dotan[b,2]

TechBetter LLC

ravit@techbetter.ai

Chapter under review for the *Handbook for Assessment in the Service of Learning*

(Editorial Team: Eleanor Armour-Thomas, Eva L. Baker, Howard Everson, Edmund W. Gordon, Steve Sireci, and Eric Tucker)

[1] Corresponding author.
[2] https://www.ravitdotan.com/

**Abstract**

Artificial intelligence (AI) creates opportunities for assessments, such as efficiencies for item generation and scoring of spoken and written responses.  At the same time, it poses risks (such as bias in AI-generated item content).  Responsible AI (RAI) practices aim to mitigate risks associated with AI.  This chapter addresses the critical role of RAI practices in achieving *test quality* (appropriateness of test score inferences), and *test equity* (fairness to all test takers) — key principles in this volume. To illustrate, the chapter presents a case study using the Duolingo English Test (DET)  – an AI-powered, high-stakes English language assessment. The chapter discusses the DET RAI standards, their development and their relationship to domain-agnostic RAI principles. Further, it provides examples of specific RAI practices, showing how these practices meaningfully address the ethical principles of validity and reliability, fairness, privacy and security, and transparency and accountability standards to ensure test equity and quality.

## 1. Introduction

*Test quality* is achieved through evidence gathering that confirms an assessment's suitability for its intended purpose. *Test equity* is attained when test scores are fair – specifically, they do not favor or disadvantage a particular group. Classical argument-based test validity theory supports test quality and equity through a chain of inferences. Inferences assume mechanisms (e.g., relevant task types) through which evidence collection ensures an appropriate test score interpretation (Chapelle et al, 2008; Kane, 1992). These include: *domain definition* (task types represent the target domain as defined), *evaluation* (test scores reflect language ability), *generalization* (test scores are reliable), *explanation* (test scores are attributable to the construct), *extrapolation* (test score are related to other language criteria), *utilization* (test scores are interpretable and meaningful for their purpose). To maintain the validity of AI-powered assessments, it is essential to evaluate AI capabilities, as they impact evidence collection, measurement, and, ultimately, test quality and equity.

With recent generative artificial intelligence (AI) advances, AI-powered assessments are becoming increasingly common. AI for assessment offers many advantages, such as automated scoring of writing and speaking, and creating larger item banks through efficiencies in automated item generation. However, there are risks. For example, bias in AI-generated item content can impact test-taker outcomes (e.g., Belzak et al., 2023; Johnson et al, 2022); this can, potentially, lead to test inequity and diminished test quality. Therefore, AI-powered assessment calls for alignment with human-centered AI values which are enacted through responsible AI guidelines and standards (von Davier A. & Burstein, to appear; Burstein, 2023; Auernhammer, 2020; The IEEE Global Initiative on Ethics of Autonomous and Intelligent Systems, 2017). Though some risks to validity are similar across traditional and AI-based assessments, AI introduces unique risks. While assessment standards address AI to some extent (American Educational Research Association (AERA), American Psychological Association (APA), & National Council on Measurement in Education (NCME) (2014) (henceforth, AERA&APA&NCME, 2014), the expanded use of AI for assessment requires more comprehensive RAI standards and practices to mitigate risks to the test validity argument that affect test score interpretation (Association of Test Publishers (ATP), 2024; US Department of Education, 2024).

This chapter illustrates how RAI assessment standards and practices help to achieve test equity and test quality of AI-powered assessments. To do so, the chapter presents a case study using the RAI standards for the Duolingo English Test (DET) – a digital-first, high-stakes English language assessment.  The chapter (i) presents the standards; (ii) explains their development; (iii) validates the DET RAI standards with ethical principles for an industry-agnostic RAI governance framework—the National Institute for Standards and Technology's (NIST) Artificial Intelligence Risk Management Framework (NIST AI RMF) (NIST, 2023); (iv) illustrates their application on the DET, demonstrating how RAI standards and practices support test equity and quality; and, (v) discusses the RAI standards implications and current known limitations for assessment.

## 2. Background and Related Work

First appearing on first-generation, computer-based assessments, AI[3]  has been used on high-stakes assessments for quite a while for automated writing evaluation  (AWE) (Lottridge et al., 2021; Shermis & Burstein, 2013, 2003; Foltz et al., 1999; Burstein et al, 1998), and speaking evaluation (ASE) (see Zechner & Evanini, 2019). Emerging more recently, *digital-first*

---

[3] Note that not all AI used for assessment is generative AI, specifically. As well, different applications use different AI systems. In this chapter, we use the term AI to refer to AI and AI-adjacent approaches, including natural language processing (NLP) and statistical modeling approaches.

*assessments* (DfA)  are designed to be administered online and leverage AI for test design (e.g., automated item generation), measurement (e.g., automated essay scoring), and security (e.g., plagiarism detection) (see Cardwell et al., 2024). DfAs are possible largely due to the availability of generative AI (OpenAI, 2023; Radford et al., 2019), allowing for automated item generation at scale (Attali et al., 2022; Khan et al., 2021).

Classical assessment validity (Chapelle et al., 2008; Kane, 2013; 1992) and fairness (Kunnan, 2000) frameworks developed for paper-and-pencil and first-generation, computer-based assessments embody the ethical principles of validity and reliability, and fairness.   These frameworks have laid the groundwork for modern assessment. However,  they do not explicitly address the use of technology for assessment.  The AERA, APA, & NCME (2014) *Standards* address automated scoring of written and spoken constructed responses, covering the scope of AI use at the time they were published.  Earlier frameworks and assessment standards also do not address aspects of AI use on tests that may impact test quality, such as the need for AI literacy training of human test developers, or broader societal issues, such as carbon emissions associated with large language model use (Faiz et al, 2024). It is not even clear how these new issues associated with AI use on assessments would be evaluated in the classical chain of inferences.

Given the growth of AI use on modern digital assessments, guidelines for RAI in learning and assessment continue to proliferate (such as ATP, 2024; US Department of Education, 2024; Organization for Economic Cooperation and Development (OECD), 2023; International Test Commission (ITC) & Association of Test Publishers (ATP) (ITC-ATP), 2022; The International Privacy Subcommittee of the ATP Security Committee, 2021; U.S. Department of Education, Office of Educational Technology, 2023). *Guidelines* make high-level recommendations for mitigating AI risks.  In  contrast to guidelines, *standards* establish principles and actionable practices to practically implement theoretical standards (such as, AERA&APA& NCME, 2014). Standards provide actionable guidance for the implementation of the ideals they present. Failing to implement the ideals creates ethical debt that may impact test equity and quality – specifically, downstream harms in the short and long run (such as hallucination in content generation; Ji et al., 2023).

RAI standards development for assessment relies on both the vast literature of assessment guidelines and standards, as well as AI ethics and governance principles. As such, RAI standards development should draw from the AI ethics literature which provides a large set of ethical principles, including fairness, transparency, explainability, privacy, security, trust, responsibility, justice, and autonomy (Memarian & Doleck, 2023; Floridi & Cowls, 2022; Fjeld et al., 2020; Jobin et al., 2019). AI ethical principles tend to be domain-agnostic, also appearing in modern AI risk frameworks (e.g., NIST, 2023). Combining domain-agnostic ethical principles promotes human responsibility to ensure equity and quality, while industry-specific standards (e.g., AERA, APA, & NCME, 2014 , 2014) and guidelines (OECD, 2023; U.S. Department of Education, Office of Educational Technology, 2023; and ITC-ATP, 2022) ensure industry-alignment.

## 3.  Case Study: The DET RAI Standards

This section presents an overview of the DET, and discusses the DET RAI standards, their development process, and their alignment with a domain-agnostic industry framework. Finally, the case study demonstrates how the systematic application of standards supports assessment quality.

### 3.1 The Duolingo English Test

The DET is a digital-first, high-stakes, computer-adaptive measure of English language proficiency, most commonly used for admissions to English-medium higher education institutions (Cardwell et al, 2024).   It assesses a test-taker's ability to use linguistic skills that are required for speaking, writing, reading and listening, as well as integrated skills associated with literacy, conversation, comprehension, and production.

The DET leverages different AI capabilities, including automated item generation, writing and speaking evaluation, and plagiarism detection for test design, measurement, and security, respectively. The DET's test-taker experience also benefits from AI. For instance, the DET's free practice test is made possible by automated item generation (i.e., to generate practice test items) and scoring (i.e., providing an instant score estimate).

The DET employs *human-in-the-loop* (HiTL) AI practices to support test quality and equity.  The DET's HiTL approach is consistent with current education and assessment policy, aiming to contribute to the test's equity and quality. Recent education and assessment policies discuss HiTL AI as crucial human oversight at critical decision points (ATP, 2024; US Department of Education, Office of Educational Technology, 2024). HiTL AI is also discussed as a real-time, human-system interaction process, whereby humans provide ongoing input to enhance AI performance (Wang, 2019).  The DET leverages human judgment and oversight for test design, measurement, and security.  For example, humans *review* automatically-generated content for test design, *label* training data to build and evaluate models for writing and speaking for measurement, and *proctor* to identify AI-generated plagiarism flags for test security. As the DET expands its use of AI to develop increasingly innovative and interactive item types, it is anticipated that HiTL practices may involve more real-time human interaction. The case study presented later provides illustrations of these practices.

**3.2 The Duolingo English Test Responsible AI Standards**

*The Four Standards*

The DET's Responsible AI standards address test equity and quality through include <u>four</u> standards that represent ethical principles aligned with the test's goals. An overview of the standards is discussed below, and more details are provided in Section 3.3.

1. The *<u>Validity and Reliability</u>* standard is crucial to ensure that the test is suitable for its intended purpose. The Validity standard evaluates construct relevance and accuracy, and the Reliability standard focuses on consistency;

2. The *<u>Fairness</u>* standard promotes democratization and social justice through increased access, accommodations, and inclusion, representative test-taker demographics, and avoiding algorithms known to contain or generate bias;

3. The *<u>Privacy and Security</u>* standard ensures (a) compliance with relevant laws and regulations governing the collection and use of test taker data; (b) ensuring test-taker privacy and (c) providing secure test administration; and,

4. The *<u>Accountability and Transparency</u>* standard aims to gain trust from stakeholders through proper governance and documentation of AI used on the test.

*Standards Development*

Five key activities led to the selection of four ethical principles used to create the DET RAI standards.

First, we conducted a literature review to examine ethical principles used for AI (including, Memarian & Doleck, 2023; Floridi & Cowls, 2022; Fjeld et al., 2020; Jobin et al., 2019). The review increased our understanding of which principles were applicable to the DET,

Second, to validate alignment between domain-agnostic, AI governance  (e.g., NIST, 2023) and assessment-specific principles, we reviewed well-recognized assessment-specific standards (AERA/APA/NCME, 2014) and guidelines (including OECD 2023; U.S. Department of Education, Office of Educational Technology, 2023; and ITC-ATP, 2022).

Third, we consulted a cross-disciplinary group of experts from computational psychometrics, language assessment, law, machine learning, and security within Duolingo, and an external RAI expert from computer science[4].

Fourth, after identifying the four ethical principles, the external RAI expert helped to articulate the rationale and overall goal of each standard, and the more detailed subgoals (i.e., practical implementation of each standard).

Finally, the standards were published as a living document that remains freely available, and open for public comment.

*Connections to NIST AI RMF*

The DET RAI Standards were validated against the independent, national, industry-agnostic NIST AI RMF (2023) *trustworthiness characteristics.*  Validation with an independent and industry-agnostic ethical framework demonstrates how our standards are aligned with prevailing best practices.

---

[4] We thank Dr. Pascale Fung for her expert guidance in the development of the DET RAI Standards (Burstein, 2023).

**Table 1.** NIST AI RMF *trustworthiness characteristics* & DET RAI Standards

| NIST AI RMF trustworthiness characteristic | Description[5] | DET RAI Standards | Description |
|---|---|---|---|
| Valid and Reliable | Ensure objective evidence, fulfilling requirements for intended use. Perform consistently in expected conditions over a period of time. | Validity and Reliability | Ensure that the test is suitable for its intended purpose. Validity standards involve evaluating construct relevance and accuracy, while Reliability standards maintain consistent performance over time. |
| Fairness with harmful bias managed | Address concerns for equality and equity by addressing issues such as harmful bias and discrimination. | Fairness | Promote democratization and social justice through increased access, accommodations, and inclusion represent test-taker demographics, and avoid algorithms known to contain or generate bias |
| Privacy Enhanced; Secure and Resilient | Adheres to privacy values such as anonymity, confidentiality  Maintain confidentiality, integrity, and availability through protection mechanisms, preventing unauthorized access | Privacy and Security | Ensure (a) compliance with relevant laws and regulations governing the collection and use of test taker data; (b) assurance of test taker privacy and (c) assurance of secure test administration. |
| Accountable & Transparent | Documents information about an AI system and its outputs for individuals interacting with the system | Accountability and Transparency | Provide thorough documentation and explanations. |

    The NIST AI RMF's trustworthiness characteristics are similar to the DET RAI standards' ethical principles in that both identify characteristics of trustworthy AI.  The NIST AI

---

[5] See NIST (2023) for complete descriptions.

RMF emphasizes the following trustworthiness characteristics: Valid and Reliable; Safe, Secure and Resilient; Accountable and Transparent; Explainable and Interpretable; and, Privacy-Enhanced, Fair – with Harmful Bias Managed.  Based on the standards development process, the DET RAI standards focus on standards which echo four of these characteristics in Table 1.

### 3.3  How the DET RAI Standards Impact Test Quality and Equity

We illustrate the application of each of the four DET RAI Standards through practices associated with their goals and subgoals. The examples demonstrate these practices uphold the standards' ideals and contribute to test quality and equity.

To illustrate, we use two DET tasks[6] — the *Interactive Reading* and *Writing Sample* tasks. We briefly describe these task types (3.3.1) See Cardwell et al. (2024) for details.   We then discuss the application of each of  the four standards in test development, measurement and security (3.3.2).

### 3.3.1. Task Descriptions: Interactive Reading and Writing Sample

The *Interactive Reading* task is a measure of a test-taker's ability to read in academic contexts. The task contains five different item types, targeting different reading sub-constructs. The item types are: Vocabulary in Context; Text Completion; Reading Comprehension; Main Idea; and Possible Title. The item response formats include multiple choice reading comprehension questions, and a question in which test takers highlight a segment of the text to respond. See Figures 2 - 6 in the Appendix for screenshots.

---

[6] The DET currently has 14 task types (Cardwell et al, 2024).

The *Writing Sample* is an independent, spontaneous writing task. Test takers receive a prompt, have 30 seconds to prepare, and then have five minutes to write their response. See Figure 7 in the Appendix for a screenshot.

### 3.3.2 Applying RAI Standards

Examples in this section illustrate how the RAI standards' practices contribute to test quality and equity.

Under the *Validity and Reliability*, and *Fairness* standards, we discuss a six-step RAI process for DET *task design*. Steps 1 - 6 are referenced throughout the discussion that follows. The six-step process also addresses aspects of measurement (e.g., scoring), and security (e.g., item exposure) issues. This is followed by a discussion about the application of the *Privacy and Security,* and *Accountability and Transparency* standards.

### 3.3.2.1 Validity and Reliability

The Validity and Reliability Standard focuses on test validity (i.e., suitability for its intended purpose) and reliability (i.e., yields consistent results), and has two main goals. The first goal aims to "specify processes required to build a validity argument*",* and the second goal aims to "evaluate AI used in test item creation, item calibration, and scoring." We illustrate four subgoals from this standard that contribute to test quality and equity.

*Develop a Description for the Test Target Domain–i.e., English Language Proficiency–to Ensure that Test Items Are Aligned with the Domain Being Measured. (Subgoal 1.1.1)*

*Rationale.* This subgoal aligns with the *domain definition* inference and involves construct definition – necessary in the design of any assessment. This is essential to later processes that use AI to automatically generate high-quality text passages.

*Implementation.*  Steps 1 and 2 illustrate the DET's task design process[7].

Step 1.  *This step articulates the target construct by human subject-matter experts (SME) from assessment science.*

For the Interactive Reading task, the target construct is <u>academic reading</u>.  This construct includes a range of reading purposes and cognitive skills categorized into two main areas (Park et al., 2022) important reading skills in university study (Grabe, 2008): *Reading for Orientation* and *Reading for Information and Argument* (Council of Europe, 2020). *Reading for Orientation* entails searching for specific information within a text and quickly understanding its general idea with limited information (Giulia Cataldo & Oakhill, 2000; Guthrie, 1988; Guthrie & Kirsch, 1987). *Reading for Information and Argument* involves understanding main ideas, learning how ideas within a text connect to each other and to the reader's prior knowledge, integrating information from multiple texts or different parts of a long text, and using the carefully curated information to interpret the text or perform other tasks (Grabe & Stoller, 2020; Head & Eisenberg, 2009; Thompson et al., 2013).

Step 2. *This step specifies a task and scoring system. This includes AI feature development and evaluation that operationalizes elements of the target construct articulated by human SMEs.*

Following the construct definition, we illustrate task specifications for passage generation for the Interactive Reading task. We create task specifications and scoring systems (including feature development and evaluation) that operationalize elements of the target construct as articulated by human SMEs.

---

[7]The remaining four steps are discussed later under subgoals 1.2.1 and 2.1.3.

Passages generated for the Interactive Reading task include two primary text categories: expository and narrative. These reflect the target language use domain: expository writing , such as in textbooks (Thompson et al., 2013; Weir et al., 2009) and news articles (Head & Eisenberg, 2009) are relevant to university students; narratives are frequently used in academic texts, such as for ethnographic reports and biographies (de Chazal, 2014). *Reading for orientation* is operationalized whereby test takers demonstrate comprehension of specific ideas (through text highlighting) and vocabulary knowledge in context (through cloze items). *Reading for information and argument* is operationalized through text completion, main idea selection, and passage title identification items.

*Evaluate AI Scoring System Accuracy and Fairness, Leveraging Human Expertise* (*Subgoal 1.1.2*)

*Rationale.* Aligned with Step 2,  it is important to evaluate AI scorers during task development to ensure that expected scoring criteria can be satisfied using computationally-derived features (e.g., grammar error detection). This is the AI analog to human scoring processes. It maps to the *explanation* inference.

*Implementation.* Implementation of this subgoal is exemplified in the Writing Sample task, where AWE is used to automatically score test-taker responses.

Human experts develop rubrics with rating criteria consistent with the Common European Framework of Reference (CEFR) levels and descriptors (Council of Europe, 2020). The rubrics define criteria based on the quality of four sub-constructs: *content* (relevance and task achievement), *discourse coherence* (organization and cohesion), *lexis* (including sophistication and correct use of vocabulary in context), and *grammar* (complexity and accuracy). These rubrics are used to train human raters who annotate (assign numerical ratings) sizable samples of essay responses using the rubric criteria – scores of 1-6, aligned with the six CEFR levels (A1-C2).  Consistent with AERA&APA&NCME (2014) standards and the *evaluation* inference, agreement rates between human raters are monitored to ensure reliable, accurate ratings (measurement). The human-rated datasets are used to compute system-human agreement as one of the key evaluation metrics used  for AI scoring model development.

*Develop (a) explainable scoring methods, and (b) interpretable AI features used for scoring that have clear alignment with domain constructs (Subgoal 1.1.3)*

 *Rationale.* Consistent with AERA&APA&NCME (2014) standards, this subgoal ensures that AI model scores are valid by requiring that scores are explainable, aligning with the *explanation* inference.  The model features discussed below measure various aspects of the domain construct and support score explanation.

 *Implementation.* The Writing Sample task is used to illustrate. The DET uses AI models to score open-ended speaking and writing responses. Drawing from the literature on NLP, linguistics, and AWE, human experts identify features to be included in the AI model. These example features represent the Writing Sample task's four sub-constructs.

- **Content:** Inverse document frequency (IDF) weighted word is used to measure similarity between a writing prompt and the test-taker response (Burstein et al., 1998; Rei & Cummins, 2016).

- **Discourse coherence**: Sentence overlap, coreference counts, and latent semantic analysis (LSA)-based sentence similarity features  (Foltz et al, 1998) similar to those implemented in the widely used Coh-Metrix (McNamara & Graesser, 2012) are used to evaluate lexical cohesion - a measure of coherence.

- **Lexis**: Proportion of words by CEFR level (Xia et al., 2016) and differential word use (DWU) (Attali, 2011) are used as measures of lexical sophistication. DWU uses outputs from n-gram classification models to differentiate low and high proficiency test-takers.

- **Grammar**: Tree depth statistics (Schwarm & Ostendorf, 2005) are used as measures of grammatical complexity. Error rates of various grammatical error types, determined through grammatical error correction and classification (Bryant et al., 2017), are used as measures of grammatical accuracy.

To make scoring models explainable, the DET uses SHAP (SHapley Additive exPlanations) (Lundberg & Lee, 2017). Response scores can be reduced to a sum of feature contributions, which can be aggregated into sub-construct contributions. Using SHAP allows for powerful scoring models with complex, non-linear relationships, such as XGBoost models (Chen et al., 2016), while providing the explainability.

*Identify AI methods for item creation, leveraging human expertise to efficiently create valid and reliable test items (Subgoal 1.2.1).*

*Rationale.* Large item banks mitigate the long-standing assessment of security issues associated with item exposure and pre-knowledge (Chen et al., 2003; LaFlair et al., 2022; Way, 1998). This subgoal manages automated item generation to efficiently create large item banks using GPT-4 (a large language model (LLM)). It aligns with the *domain definition* inference in that it attends to construct relevance.

*Implementation.* For illustration, we switch back to the Interactive Reading task type, since it involves a more involved item generation process than the Writing Sample item type. Steps 3-4 below discuss the implementation.

Step 3. *This step develops prompts that elicit content and questions from a large language model (LLM) (GPT-4 for the DET) that meet the specifications for DET tasks. Note that the prompts and specifications are developed by SMEs who are machine learning experts.*

To generate content and texts at scale for the Interactive Reading task, machine learning and assessment scientists collaborate to develop prompts that align with the task specifications (See subgoal 1.1.1).

Step 4. *This step uses GPT-4 for large-scale generation of content and tasks.*

The text types for Interactive Reading (expository and narrative) can be generated at scale via the prompting process. One approach to this is to employ in-context learning (Dong et al., 2022), where exemplar texts are provided as part of the prompt submitted to the LLM. In this approach, prototypes of narrative and expository readings are shown to the LLM and then reading passage outputs are generated. The outputs could be conditioned on any relevant target characteristic, such as a list of STEM subjects or topics.

For main idea and possible title item types (mentioned earlier), potential answers are generated and evaluated based on similarity to the passage. For comprehension questions, the model generates questions and answers, filtering out those with undesirable characteristics (such as, extreme lengths, or poor alignment with the passage). For the text completion item type, candidate target sentences are identified based on the probabilistic likelihood of their occurrence in the text. To generate distractors for main idea and possible title items, alternative texts and questions are generated and their main ideas and titles are used as incorrect answers. Candidate distractors are selected based on metrics such as vector embedding similarity and LLM log probability. For vocabulary in context items, a different process is used: words for deletion are selected based on likelihood, rank order, syntactic and semantic information, and distance from other elided words. Candidate distractors for deleted words are then selected from the model's likelihood ranking (targeting lower likelihood values) for the candidate words (Attali et al., 2022). With regard to quality, human review of generated content and items is discussed later as part of Step 5.

### 3.3.2 Fairness

The Fairness Standard addresses test equity explicitly. It aims to ensure that test takers have equal opportunity to succeed and that AI is free of algorithmic bias. It consists of two main goals. The first goal aims to "specify how the use of AI facilitates test-taker access, accessibility, and inclusion," and the second goal aims to " specify test-taker demographic representation, and algorithms known to contain or generate bias." We focus on two subgoals that address fairness and bias (FAB) review of items, and the mitigation of algorithmic bias.

*Develop and apply fairness and bias item review principles for inclusion that eliminate construct-irrelevant barriers and ensure that cultural and linguistic factors do not impede accessibility and inclusion* (*Subgoal 2.1.5)*

*Rationale.* By focusing on access, accessibility, and inclusion, this standard aims to create a more equitable testing environment for individuals from diverse backgrounds and with varying needs. One of the ways this is achieved is developing and applying item reviews to increase inclusion and eliminate potential biases in automatically-generated test content. This aligns with the *explanation* inference in that the review manages task (passage) characteristics.

*Implementation.* Humans review the content and tasks to identify sensitive content and low quality items. The human review process improves the item design process and the prompt development based on the human review and feedback. These processes are achieved through Steps 5-6.

Step 5. *This step requires that we review the content and tasks by humans.*

To mitigate the potentially negative impact of poorly constructed or distracting content from automatically-generated items, we remove that content through a human review that includes a *fairness and bias* (FAB) review and an item quality (IQR) review. Such reviews are a long-standing tradition in test development (AERA/APA/NCME, 2014). FAB review ensures that items are factually accurate, and do not contain content that may introduce culturally sensitive or inaccessible topics that might upset or distract the test-taker and introduce construct-irrelevant variance. What is different about human review of AI-generated content is required awareness about potential issues specifically associated with LLM outputs which are unlikely to occur with human-created items, such as LLM hallucinations. Item quality review evaluates the content and questions to ensure that they do indeed sufficiently represent the relevant text types (narrative and expository) and sub-constructs targeted by the questions.

For both types of reviews, reviewers are trained using in-house materials. Item quality reviews are tailored to each task type. Where relevant, they are informed by state-of-the-art item writing guidelines (Haladyna et al., 2002). The FAB guidelines are an expanded version of Zieky (2015); they are tailored by the DET test developers through regular discussions to avoid inclusion of sensitive content. Additionally, DET test developers survey test takers to understand what types of content they would like to read when taking the test, allowing for test-taker input during test design. While these test-taker surveys do not create fully democratic involvement in the test design process (Jin, 2023; Shohamy, 2001), they incorporate for test taker input about the test content.

Step 6. *This step aims to improve the item design process and the prompt development based on the human review and feedback.*

Here, we close the feedback loop between content generation and feedback that is collected from reviewers, weekly research discussions, and test-taker content surveys responses. Information gathered from these sources is used to improve item generation procedures. The information collected through the reviews can be used to improve the prompts for the LLMs, our automated filters of generated content, and our FAB and IQR reviewing guidelines.

This six-step process: 1) aligns the AI-assisted item development process with traditional approaches (e.g., construct articulation, development of specifications, content review) and 2) introduces human evaluation of AI outputs (e.g., content generation and automated scoring) which are likely to have characteristics unique to LLMs.

*Evaluate and document demographic representation in data sets used to build AI. Documentation should describe how representative (inclusive) the data are with regard to DET test takers (Subgoal 2.2.1)*

*Rationale.* It is crucial to document and evaluate demographic representation in datasets used for AI-powered assessments. This aids in creating inclusive data sets that represent the test-taker population, and aims to mitigate biases in test-taker outcomes, downstream. This is aligned with the *evaluation* inference as it builds in the awareness of demographic groups.

*Implementation.* An important part of developing DET's automated writing scorer for the Writing Sample task type is curating human-rated datasets used to train and evaluate scoring models. When building these datasets, Writing Sample responses are sampled to include a roughly equal number of men and women from seven different L1[8] backgrounds that are most common among the test-taker population and cover a broad range of language families: Arabic, Mandarin Chinese, Telugu, English, Spanish, Gujarati, and Bengali. This ensures that the model is trained and evaluated on a diverse range of L1 backgrounds, promoting measurement quality.

---

[8] L1 refers to the test-takers' self-reported first language.

*Evaluate and document bias associated with automatically-generated item content (e.g., Fairness and Bias Review Guidelines), and proficiency measurement* (*Subgoal 2.2.3*)

*Rationale.* It is essential to evaluate and document known algorithmic bias in AI used in assessment processes, such as test security, design, and measurement. This includes managing potential bias associated with automatically-generated item content and proficiency measurement. This is aligned with the *evaluation* inference.

*Implementation.* The DET implements this subgoal for proficiency measurement (scoring) by using Differential Rater Functioning (DRF) analysis  (Jin & Eckes, 2021; Myford & Wolfe, 2004) on all scorers for open-ended writing and speaking tasks, including the Writing Sample task.  Specifically, these scoring models are evaluated on the representative dataset (mentioned in subgoal 2.2.1) to quantify any bias they may have with respect to sensitive background characteristics (e.g., gender or L1) after controlling response quality. We perform this kind of analysis at both the feature and score level to identify potential differential performance test-takers groups. A similar analysis called differential item functioning (DIF) is used to detect bias at the item level (Holland & Wainer, 2012) caused by automatic item generation. The DET conducts differential item functioning (DIF) on its item bank (Belzak et al., 2023), flags items with potential bias, and sends them back for FAB review.

### 3.3.1.3 Privacy and Security

The Privacy and Security Standard seeks to ensure that the test administration process is secure, fair, and reliable, while protecting test-taker privacy and preventing cheating. This standard consists of three goals. The first goal aims to "specify methods to ensure privacy and security associated with data origin, data collection and processing, and data management". The second goal aims to "specify how to maintain test-taker privacy, item security, and test-taker security during test administration". The third goal aims to "specify fair and reliable test security proctoring protocols, item pool development, and psychometric procedures for test security." In this section, we highlight a subgoal from this third goal.

*Define, document, and implement human-in-the-loop AI proctoring protocols that fairly and reliably identify novel and known cheating behaviors (Subgoal 3.3.1)*

*Rationale.* This subgoal focuses on ways to use AI to identify cheating behaviors, and develop protocols. It supports DET proctors'[9] use of AI-enabled tools to make informed, equitable decisions about cheating behaviors that show up on high-stakes assessments. For instance, test takers might hire others to help them test, use texts written by others (traditional plagiarism), and, most recently, use AI tools, such as LLMs (e.g., GPT) to generate responses (Khalil & Er, 2023). This is aligned with the *evaluation* inference.

*Implementation.* We illustrate the implementation of this subgoal on traditional plagiarism. Traditional plagiarism is a known problem on high-stakes language assessments (Wang et al., 2019). For example, test takers will memorize long, generic essay responses that they superficially adapt to respond to a writing prompt. Such cases can often be detected using

---

[9] Note that the DET employs asynchronous proctoring (see Duolingo English Test, 2021)

AI models with features that quantify overlap between texts (Foltýnek et al., 2020). It is important to distinguish between deceitful plagiarism and benign text overlap (Chandrasoma et al., 2004; Pecorari & Petrić, 2014).

During proctoring, the DET uses AI to compare test-taker writing responses to a database of relevant Internet content and writing responses from historical test sessions. Matches are flagged and shown to proctors (Figure 1 in the Appendix). The DET's plagiarism tool shows the sources where matches were found, and highlights the overlapping text. This demonstrates human-in-the-loop AI, as AI tools help human proctors with decision-making. (Note that subgoals within the Accountability and Transparency standard address *AI literacy* requirements to ensure that proctors understand how AI tools work.)

### 3.3.1.4 Accountability and Transparency

The Accountability and Transparency Standard seeks to build trust with stakeholders. The standard is satisfied through six goals related to the DET's documentation and dissemination about AI use. (See Burstein, 2023 for details about the six goals). We illustrate standards' application with the  first (4.1), second (4.2) and fifth (4.5) goals. The <u>first</u> goal  is to "assess how AI processes impact stakeholders".  The <u>second</u> goal is related to documenting how "AI is used for building the validity argument, test item creation, test item calibration, and scoring". The <u>fifth</u> goal focuses on "disseminating research about use of AI to various stakeholder communities".  Arguably, goals in this standard align with the *utilization* inference, since the documentation offers stakeholders explanation about different aspects of test design, measurement and security.

*Document external factors that result in a need to modify AI (Subgoal 4.1.3)*

*Rationale.*  External factors can affect the impact of AI used on an assessment.  For example, changes in test-taker population may increase bias in the form of differential item functioning (DIF) or differential rater functioning (DRF).

*Implementation.*  The DET's Analytics for Quality Assurance for Assessment (AQuAA; Liao et al., 2022) system addresses this subgoal. The AQuAA system provides weekly reports on several relevant metrics that reflect the quality and comparability of the test scores over time, particularly with respect to shifts in test-taker demographics.

*Document AI used for building the validity argument, test item creation, test item calibration, and scoring (Goal 4.2)*

*Rationale.* This documentation ensures that internal stakeholders fully understand how AI is being used to ensure that it is being used appropriately to support a fair, valid, and reliable test.

*Implementation.* To help satisfy this goal and all its subgoals, the DET documents and controls the use of AI through its Exam Change Proposal (ECP) process. For example, when a new scoring model is developed for task types, such as the Writing Sample, the evidence for the scoring model's validity, reliability, and fairness is collected and documented in an ECP document. Similarly, when a new item is developed for operational use, the evidence that supports the launch of the item is documented in an ECP. Before proposed changes are implemented, the document is reviewed and approved by multiple experts, including DET senior assessment and AI researchers.

*Disseminate research about use of AI to various stakeholder communities (Goal 4.5)*

*Rationale.* Outcomes from high-stakes assessments can profoundly impact test-takers' educational goals. Test developers should clearly communicate with stakeholder communities about how AI is used across the assessment ecosystem for test design, measurement, and security.

*Implementation.* To reach different stakeholder audiences, DET researchers regularly disseminate research in the form of blog posts, white papers, and peer-reviewed articles. For example the launch of the Interactive Reading task was accompanied by a white paper describing the task and what it measures (Park et al., 2022), and a subsequent technical, peer-reviewed article describing the procedures for automated generation of the task (Attali et al., 2022).

## 5. Limitations and Future Work

Organizational RAI guidelines and standards are not a one-time exercise (PwC, 2024). Organizations that build and deploy AI-powered assessments should commit to integrating RAI into the full assessment ecosystem, as a test is developed and deployed.  As well standards development should  be updated in step with AI use on the test, and the evolving understanding of its risks which can impact test equity and quality. Known limitations for the DET RAI standards are discussed here.

*AI advances.* GPT-4o was released (OpenAI, 2024) only slightly ahead of the time this chapter was being written. GPT-4o is a much more powerful LLM than had previously existed. Its multimodal generation capabilities creates opportunities, such as use for innovative item types. At the same time, it presents risks (such as deep fakes which have implications for test security). To maintain test equity and quality, assessment developers need to consider how this new technology can be responsibly used for assessment.

*Fairness.*  Fairness issues span across all standards. For example, the use of AI to detect traditional plagiarism[10] was described in the Privacy and Security standard. Since it is acknowledged that AI exhibits biases, it is possible that these detectors introduce biases into the plagiarism evaluation. Given the pervasive nature of fairness issues, one approach to consider is making fairness a cross-standard narrative, decreasing the likelihood that fairness issues fall between the cracks.

---

[10]The DET recently introduced methods to detect plagiarism behaviors associated with the use of LLMs, and a manuscript describing the methods is in preparation.

*Additional RAI Standards.*  The DET RAI includes four standards: Validity and Reliability, Fairness, Privacy and Security, and Accountability and Transparency. These topics were chosen through a process of literature review, consultation with experts, and internal deliberation. However, naturally, some topics were left out.  For example, two important issues the DET RAI Standards do not cover are environmental and labor impacts. The DET's environmental impacts include the carbon emissions and water consumption of the generative AI applications involved in the assessment process, as these impacts are notoriously high (Strubell, Ganesh & McCallum, 2023).  The  DET is working on estimating the environmental impact of the AI models used on the test. Such impact also includes substitution effects associated with test takers taking the DET instead of an alternative English language proficiency test. For example, if each DET test session were instead replaced by a physical, in-person test session at the closest test center, we estimate that it would require approximately tens of millions of additional kilometers in travel each year.  Labor impacts could affect Duolingo's employees as a result of the adoption of generative AI. Currently, there have been no negative impacts as the company's employees have been retrained to integrate generative AI assistance.

The DET's  approach to these and other important AI ethics issues is incremental, aiming to increase the scope of the DET RAI standards over time.

## 5. Implications & Conclusion

Intended implications of this chapter were to increase AI responsibility in assessment with attention to how it may impact test quality and equity. To do this, we provided a case study that showcases RAI standards customized for assessment; explains the standards' development process; validates the standards against an AI industry standard – i.e. the NIST AI RMF trustworthiness characteristics; illustrates the standards' implementation; and, facilitates an opportunity for critical professional and public engagement.

The DET RAI standards illustrate one example of how RAI standards and practices can be developed and applied for assessment. Through concrete examples of standards application, this chapter demonstrates how RAI standards contribute to test quality and equity, and ensure that test score interpretations are trustworthy and appropriate. The broader assessment community is invited to consider the DET RAI standards if they choose to develop standards for other assessments. As well, the DET RAI standards can be used to inform professional assessment standards addressing RAI.

## Acknowledgements

We are grateful to the anonymous reviewers for their insightful comments. Many thanks to our Duolingo English Test colleagues: Mancy Liao, for providing content for plagiarism detection; and, Ed Fu, for content related to environmental impact.

# Appendix

**Figure 1.** Screenshot of the proctor interface.

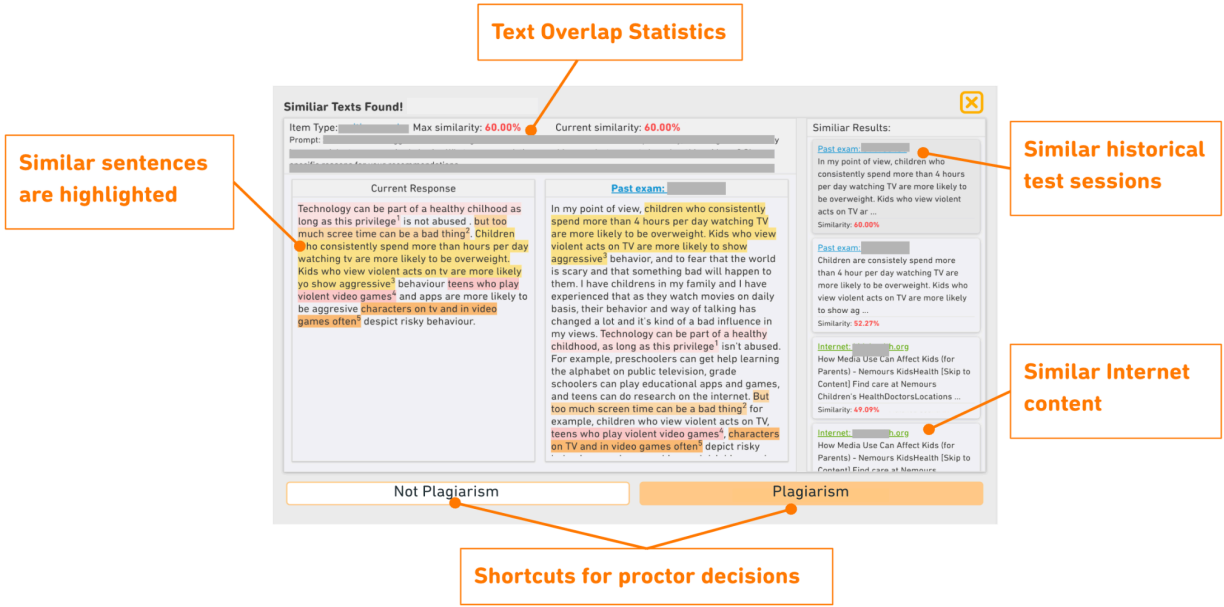

**Figure 2.** *Interactive Reading: Vocabulary in Context (Cloze)*

**6:53** for the next 6 questions

PASSAGE

Biophysics is the study of the physical properties of living things. This field refers to physics, which [1] the science of matter and energy, and also to biology, the science of living [2]. Biophysicists study the physical [3] of organisms and the [4] of physical processes on [5] things. For example, biophysicists might study the effect certain chemicals [6] on living cells, determine how tiny structures within cells work, or explain how injuries and diseases [7] the structure of skin. Some biophysicists also [8] the interaction of radiation with [9] systems.

**Select the best option for each missing word.**

1. Select a word
2. Select a word
3. Select a word
4. Select a word
5. Select a word
6. Select a word
7. Select a word
8. Select a word
9. Select a word

NEXT

**Figure 3.** *Interactive Reading: Text Completion*

**5:38** for the next 5 questions

PASSAGE

Biophysics is the study of the physical properties of living things. This field refers to physics, which is the science of matter and energy, and also to biology, the science of living things. Biophysicists study the physical properties of organisms and the effects of physical processes on living things. For example, biophysicists might study the effect certain chemicals have on living cells, determine how tiny structures within cells work, or explain how injuries and diseases affect the structure of skin. Some biophysicists also study the interaction of radiation with biological systems.

Biophysics is an interdisciplinary field; it is not limited to physics or biology. Biophysicists might also work on projects involving chemistry, geology, and other fields.

**Select the best sentence to fill in the blank in the passage.**

○ They have even studied the physical properties of the cells in the human body.

○ Biophysics is an interdisciplinary field; it is not limited to physics or biology.

○ Forensic science is the application of the techniques of the physical sciences to analyze evidence.

○ The discovery of quantum mechanics in 1925 ushered in a new world of physics.

NEXT

**Figure 4.** *Interactive Reading: Comprehension Questions*

**4:21**  for the next 4 questions

PASSAGE

Biophysics is the study of the physical properties of living things. This field refers to physics, which is the science of matter and energy, and also to biology, the science of living things. Biophysicists study the physical properties of organisms and the effects of physical processes on living things. For example, biophysicists might study the effect certain chemicals have on living cells, determine how tiny structures within cells work, or explain how injuries and diseases affect the structure of skin. Some biophysicists also study the interaction of radiation with biological systems. Biophysics is an interdisciplinary field; it is not limited to physics or biology. Biophysicists might also work on projects involving chemistry, geology, and other fields.

**Click and drag to highlight the answer to the question below.**

**How does biophysics relate to physics and biology?**

Highlight text in the passage to set an answer

NEXT

**Figure 5.** *Interactive Reading: Main Idea*

**3:46** for the next 3 questions

PASSAGE

Biophysics is the study of the physical properties of living things. This field refers to physics, which is the science of matter and energy, and also to biology, the science of living things. Biophysicists study the physical properties of organisms and the effects of physical processes on living things. For example, biophysicists might study the effect certain chemicals have on living cells, determine how tiny structures within cells work, or explain how injuries and diseases affect the structure of skin. Some biophysicists also study the interaction of radiation with biological systems. Biophysics is an interdisciplinary field; it is not limited to physics or biology. Biophysicists might also work on projects involving chemistry, geology, and other fields.

**Select the idea that is expressed in the passage.**

○ Biophysicists study the physical properties of organisms and how they interact with their environments.

○ Electric charges can cause molecular reactions by changing their shape, size, or position.

○ Living things are always in motion and they use this motion to perform many functions.

○ Cells and tissues are the basic building blocks of living things, such as humans and animals.

NEXT

**Figure 6.** *Interactive Reading: Possible Title*

PASSAGE

Biophysics is the study of the physical properties of living things. This field refers to physics, which is the science of matter and energy, and also to biology, the science of living things. Biophysicists study the physical properties of organisms and the effects of physical processes on living things. For example, biophysicists might study the effect certain chemicals have on living cells, determine how tiny structures within cells work, or explain how injuries and diseases affect the structure of skin. Some biophysicists also study the interaction of radiation with biological systems. Biophysics is an interdisciplinary field; it is not limited to physics or biology. Biophysicists might also work on projects involving chemistry, geology, and other fields.

**Select the best title for the passage.**

○ Computer Simulation of Living Systems

○ The Nature of Motion

○ The Processes of Life

○ An Introduction to Biophysics

NEXT

**Figure 7.** *Writing Sample*

**4:55**

## Write about the topic below for 5 minutes.

Describe behaviors that are important for success in school. Why are these behaviors important? How would some of these behaviors help you? Use examples from personal experience and observations to explain your perspective.

Your response

CONTINUE AFTER 3 MINUTES